# Current Mapping of Amorphous LaAlO$_3$/SrTiO$_3$ near the Metal-Insulator Transition

Anders V. Bjørlig,[1] Dennis V. Christensen,[2] Ricci Erlandsen,[2] Nini Pryds[2] and Beena Kalisky[1]*

1. Department of Physics and Institute of Nanotechnology and Advanced Materials, Bar-Ilan University, Ramat Gan 5290002, Israel

2. Department of Energy Conversion and Storage, Technical University of Denmark, Fysikvej 8 310, 2800 Kgs. Lyngby, Denmark

*Email: beena@biu.ac.il

**Abstract**

The two-dimensional electron system found between LaAlO$_3$ and SrTiO$_3$ hosts a variety of physical phenomena that can be tuned through external stimuli. This allows for electronic devices controlling magnetism, spin-orbit coupling, and superconductivity. Controlling the electron density by varying donor concentrations and using electrostatic gating are convenient handles to modify the electronic properties, but the impact on the microscopic scale, particularly of the former, remains underexplored. Here, we image the current distribution at 4.2 K in amorphous-LaAlO$_3$/SrTiO$_3$ using scanning superconducting-quantum-interference-device microscopy while changing the carrier density *in situ* using electrostatic gating and oxygen annealing. We show how potential disorder affects the current and how homogeneous 2D flow evolves into several parallel conducting channels when approaching the metal-to-insulator transition. We link this to ferroelastic domains and oxygen vacancies. This has important consequences for micro- and nanoscale devices with low carrier density and fundamental studies on quantum effects in oxides.

**Introduction**

Since the discovery of the two-dimensional electron system (2DES) at the interface between the band insulators LaAlO$_3$ (LAO) and SrTiO$_3$ (STO), a variety of fascinating properties have been found which are not observed in the single compounds. The interface exhibits metallic conductivity with high mobility,[1] coexisting magnetism and superconductivity,[2] ferroelectricity[3], and strong spin-orbit coupling.[4,5] The majority of these phases are highly sensitive to changes in the carrier density at the interface.[6] The superconducting phase in LAO/STO with $T_c \approx$ 200 mK can be turned insulating through a quantum phase transition,[7,8] and a new gate-tunable metallic phase where electrons are paired together without showing superconductivity was further found.[9,10] Similarly, the spin-orbit coupling can be tuned by varying the carrier density, as it is tied to the electron occupancy, with a strong coupling found for an occupancy close to the Lifshitz transition.[4]

Local measurements of the electronic properties have contributed to understanding the interface properties in STO-based heterostructures. Such investigations can be done by using local tools such as scanning superconducting-quantum-interference-devices (SQUID)[11,12] or scanning single-electron-transistors (SET),[13,14] which generate images of the magnetic flux or electrostatic potential with typical spatial resolutions of ~1 µm and ~600 nm, respectively.[3] Such local studies have revealed the role of the crystal structure in modulating the electronic properties of the interface below 105 K, where STO becomes tetragonal through a ferroelastic transition.[15] The ferroelastic domains can be manipulated by using electric fields[3] and temperature cycling.[16] Scanning SQUID revealed a modulated current flow over the domain pattern, which lead to anisotropy in the electronic properties depending on the domain configuration.[16,17] When electrostatic gating is used to bring the system close to the metal-insulator-transition, this modulation over the domain pattern leads to a nonuniversal behavior.[18] Consequently, the domain configuration can play a crucial role, particularly in mesoscopic devices where the length scale of the domain patterns is comparable to the characteristic lengths of the device.

The carrier density is controlled dynamically by electrostatic gating (using a dielectric material or an ionic liquid[19]), statically determined by variations in the donor concentrations (such as La or Nb for STO) or tuned during fabrication.[20] One way to dynamically tune the carrier density in oxide materials is to modulate mobile oxygen vacancy donors by annealing.[21,22] This has been investigated in heterostructures where amorphous-LaAlO$_3$ (a-LAO) is deposited on STO. Here the formation of the 2DES happens during the growth of the top layer by the formation of oxygen vacancies, which act as electron donors.[23–25] By careful annealing of the a-LAO/STO in an oxygen environment, the carrier density can be controlled over a wide range, leading to a conducting interface spanning metallic to insulating.[6] This has been used in conjunction with electrostatically induced carrier density changes, where annealing was used to raise the resistance of STO-based interfaces to a level where nanoscale conducting lines could be induced by using a conducting atomic force microscopy tip.[26] However, it remains unclear how the local electronic properties are affected by the modulations of the carrier density and the local crystalline order.

Here, we modulate the carrier density of a-LAO/STO *in situ* by means of annealing and electrostatic gating while locally imaging the resulting current distribution using scanning SQUID. We studied the transition from metallic conductivity to an insulator phase for both tuning methods: annealing and electrostatic gating. We show that in both cases the current distribution evolves from homogeneous 2D conduction to several discrete percolative paths close to the insulator.

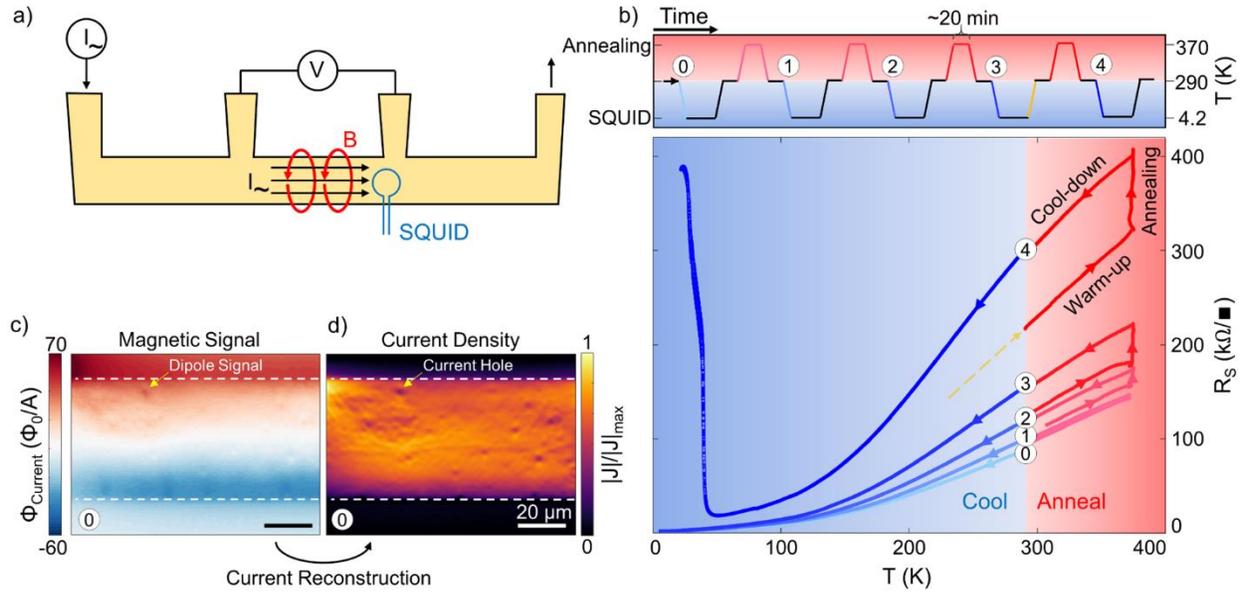

**Figure 1.** (a) Schematic of the 4-point device used in the study and the SQUID measurement of the magnetic fields (red) from the current flow (black). (b) Top: the a-LAO/STO device was cooled five times starting from unannealed (0) to insulating (4). Bottom: sheet resistance measurements of all annealing (right, red) and cooling (left, blue) steps. The dotted orange line indicates the intersection of the warm-up curve after the fourth cool-down, and the arrows on the data curves show the direction of the measurement. (c) Magnetic flux generated by current flow as recorded by the SQUID scaled by the applied current prior to annealing. White lines indicate the borders of the device defined as 100 μm wide. (d) By use of Fourier analysis, the magnetic flux from (c) can be used to reconstruct the current distribution in the device revealing several small holes in the current distribution.

**Results**

We used a scanning SQUID with a ~1 μ$\Phi_0$ magnetic field sensitivity to record the magnetic flux through a 1 μm wide pick-up loop, which we scanned near the sample surface (see the Methods section). To study the evolution of the current distribution, we fabricated a 100 μm wide channel with four voltage probes (Figure 1a) from amorphous LAO grown at room temperature on $TiO_2$-terminated STO using pulsed laser deposition (PLD) (see the Methods section). To anneal the sample *in* situ we equipped the scanning SQUID setup with a heater, atmosphere control, and a metallic back-gate. The sheet resistance of the a-LAO/STO device was monitored by using DC transport measurements during cool-downs and annealing stages (Figure 1b). The annealing of the device was performed in an atmosphere of a 4:1 nitrogen:oxygen mixture. As the device was warmed up, the resistance increased with temperature, primarily as a result of increased scattering with longitudinal optical phonons.[26,27] The temperature was then kept at 370 K for ~20 min, which caused the resistance to further increase as oxygen was reintroduced to the crystal.[21] After cool-down to room temperature following four consecutive annealings, the room temperature resistance increased from 88 to 300 kΩ/□, corresponding to a drop in carrier density from $7.1 \times 10^{12}$ to $2.1 \times 10^{12}$ cm$^{-2}$ (see the Methods section). After each annealing step the sample was cooled to 4.2 K for SQUID measurements, and after the fourth annealing step the resistance showed insulating behavior. The resistance values at 4.2 K after the first three annealing steps showed a small increase from 2.3 to 2.6 kΩ/□, corresponding to an increase of the residual resistance ratio ($R_s$(300 K) / $R_s$(4 K)) from 38 to 115. We attribute this change to an increase in electron mobility at low temperature after gentle annealing, similar to reports from previous studies in STO-based heterostructures.[28,29] The warm-up curves recorded after SQUID imaging at different back-gate voltages showed an increased resistance compared to the cool-down curves and several characteristic signatures at specific temperatures (Figure S2). This behavior is consistent with previous reports in LAO/STO where it was related to thermally assisted detrapping of carriers,[30,31] presumably after trapping the carriers by application of a back-gate voltage at 4.2 K.[30,32] Before annealing, we applied an alternating current to the device and imaged the associated magnetic flux with the SQUID at 4.2 K (Figure 1c). Inside the conducting channel the magnetic flux gradually changes between a maximum and minimum located near the two opposite edges, while it goes to zero outside. The flux is further modified by several small dipole-like patches. We use Fourier analysis to reconstruct the measured magnetic flux into a map of the current density in the device[33,34] (Figure 1d). Here, we can see that the patches in the magnetic flux represent regions of reduced current density. These "current holes" have been observed previously in various amounts in crystalline LAO/STO.[16,18]

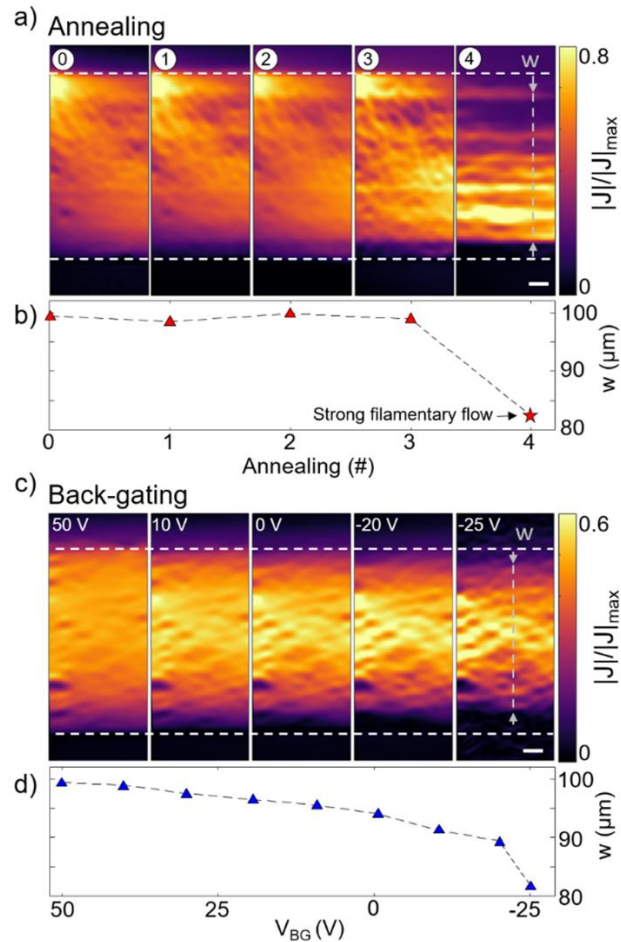

**Figure 2.** (a) Current distribution images for the annealed device after each cool-down in a 50 μm section. Here, the current transitions from 2D flow to discrete channel flow. The increase in current density in the top left corner stems from the corner of the sample where current is higher due to focusing. White lines indicate the initial borders of the device. The scale bar is 10 μm. (b) Effective width at each annealing step, defined as the distance between the two outermost points of current flow at the gray line indicated in (a). After the final annealing step, the flow is filamentary rather than homogeneous, and the effective width is not well-defined (star). Here, we have extracted the distance between the bottom edge and the topmost current-carrying channel. (c) Current distribution images as the carrier

density is reduced by a back-gate voltage. The scale bar is 10 μm. Here, the width is continuously reduced due to the size difference between the back-gate and channel. At -25 V the homogeneous current flow has evolved into several channels that are interrupted by current holes. (d) Effective width of the back-gated channel (c) defined similarly as in (b) displaying the effect of the back-gate on current distribution.

We then imaged the evolution of the current flow after each annealing (Figure 2a). The first two annealing steps did not result in substantial changes to the current distribution compared to the as-deposited measurements. In particular, the position of the current holes in the images remained static. After the third annealing step, more visible changes appeared. We observed a stronger contrast between the current-carrying parts of the device and the obstructing holes, as current is forced away from the holes and focused in the remaining regions. The locations of the holes remained the same, but regions with an overall homogeneous current flow also became distorted. At the lower edge of the device, a section of the channel is close to the insulating state. After the fourth annealing step, we found that most of the device had become insulating while the remaining current runs in channels aligned with the direction of the current flow. The channel structure was already weakly visible in the previous annealing steps, but it only became clearly visible after the carrier density was strongly reduced. In this experiment, the location of the channels did not change through thermal cycling and annealing. The configuration of channels is determined by the strain profile of each sample.

Next we extracted the effective width of the device at each annealing step, i.e, the width of the current-carrying parts of the channel (Figure 2b). One of the inherent properties when tuning devices with back-gates is an inhomogeneous response across the device due to the geometry of the capacitor: a finite top plate and an "infinite" bottom plate.[35] This geometry leads to a different voltage on the edge of the tuned device compared to the center, leading to earlier depletion at the edges. Hence, there is a continuous change to the effective width of the device as carriers are removed by the gate. Therefore, the effective width is a property that differentiates between annealing and back-gating, as ways to reduce the carriers. The extracted effective widths in Figure 2b show that annealing does not lead to a continuous change, but rather to a dramatic reduction of the width (~20%) near the transition to an insulator (Figure 2a), where the flow becomes filamentary.

We now compare the spatial distribution of the current flow tuned by annealing to tuning by electrostatic gating. We imaged the current distribution of an as-deposited sample at 4.2 K as a function of back-gating (Figure 2c). These images are taken after the initial back-gate sweeps to avoid hysteresis effects caused by charge trapping.[30] As the carrier density is electrostatically reduced, we observe an increasing contrast in the images, similar to our observations during the annealing procedure. However, the picture is now controlled by two factors: (1) the limitation of the current flow by current holes that expel more current and force it to move in the remaining conducting regions and (2) the narrowing of the effective width of the channel caused by the application of a negative back-gate voltage (Figure 2d). This forces the current toward the center of the device. The annealing process was mostly dominated by (1).

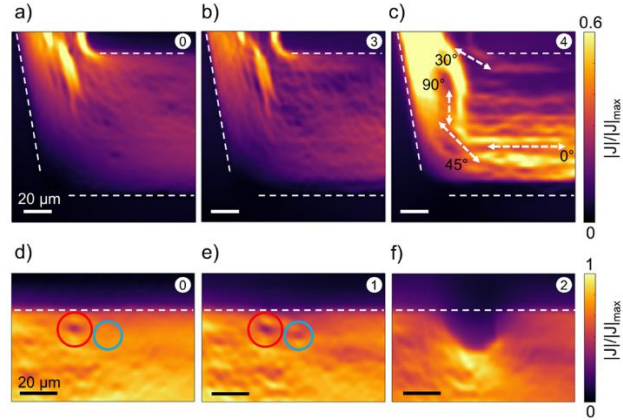

**Figure 3.** (a-c) Current distribution at the corner of the device with annealing. After the final annealing we observe not only channels following the three crystallographic orientations but also different orientations highlighted by the 30° channel. (d-f) Current density image of an evolving disorder potential near the edge of the channel. First we observe an increase in size of a current hole (red) and the appearance of a new hole (light blue) from (d) to (e). In (f) the holes combine into a larger insulating area in the 2DES, locally reducing the width of the channel.

The modulation of current flow over channel-like features observed in Figure 2a for amorphous LAO/STO shares similarities with the modulated flow observed for crystalline LAO/STO, attributed to the tetragonal domains in STO.[16] This may lead to a new configuration of channels after cycling the temperature above 105 K. In our case the channels do not show dramatic changes after cycling and annealing. The configuration is determined as the sample cools through the structural transition (105 K), and the local strain profile is a dominant factor in its formation. Another supporting argument to the role of the STO domain structure comes from examining the orientation of the channels. Ferroelastic domains in STO should result in channels oriented along the [100], [010], and [110] crystallographic directions. To investigate this, we studied the evolution of the current distribution near one of the corners of the device at various annealing steps (Figure 3a-c). In this area, we observe modulation occurring along the three crystallographic directions combined with the presence of regions where the channel-like flow is curved or not flowing along these three primary crystallographic directions. Many of the channels are also perturbed by current holes, such that the current trajectories wiggle rather than following straight lines. While the locations of the insulating holes, in general, remain static, some regions undergo the following changes. The as-deposited sample shows an overall homogeneous current distribution only perturbed by holes (Figure 3d). However, subsequent annealing steps reveal new insulating holes in the current flow (blue circle in Figure 3d-e), while some holes display a small increase in size (red circle in Figure 3d-e). In some regions, we observe that large insulating patches are formed (Figure 3f) and evolve (Figure S3) with annealing. This change in the conduction landscape severely redistributes the current flow in the remaining conductive parts of the device.

Our observations show that oxygen annealing results in the emergence of new current holes, while the existing holes are immobile up to 370 K and grow as the vacancies are annihilated rather than disappear. Based on this, oxygen vacancies alone are unlikely the origin of the current holes, and other defects in the crystal structure should be considered.

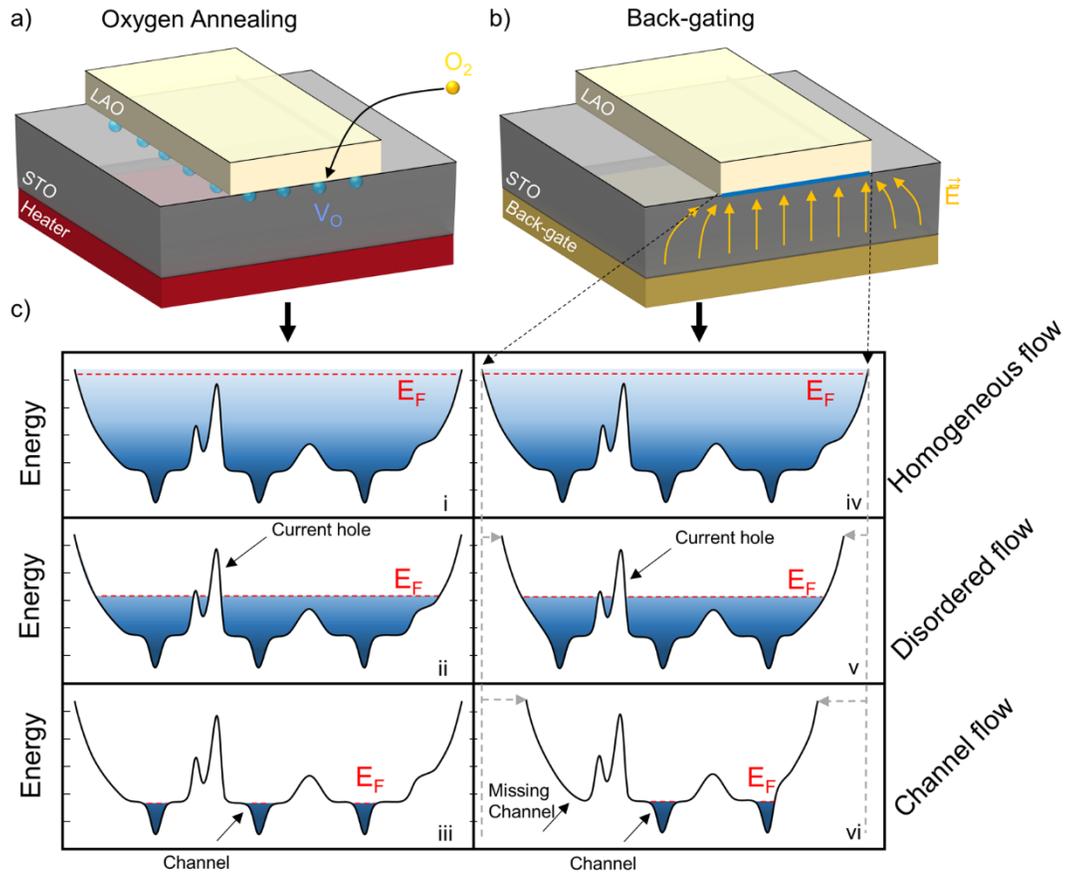

**Figure 4.** (a-b) Concept schematics of oxygen annealing and back-gating as paths to reduce the carrier density in a-LAO/STO. In (a) the sample is heated in an oxygen-rich environment which allows reintroduction of oxygen into the crystal lattice resulting in less free carriers. In (b) carriers are removed from the 2DES through voltage applied to a nearby electrode. (c) We understand the current distribution images in the framework of a disordered potential landscape inside the channel for annealing (i-iii) and back-gating (iv-vi). As the carrier density is reduced more of the disorder potential is revealed, resulting in channel-like current flow close to the insulating phase. In the gating case, the width reduction removes one of the conducting channels. For the annealing case, note that the location of wells may change due to a new configuration of domain structure formed during the temperature cycle.

Overall, the current flow has the highest degree of homogeneity in annealing or gating conditions where the conductivity is high. At low conductivity, the spatial modulations become a dominant feature due to a combination of the influence of domains and of the disordered landscape. These observations can be explained by considering a disordered potential well inside the device with smaller narrow wells representing the ferroelastic domain structure for three different levels of carrier densities (Figure 4). Initially, the carrier density is high, and the Fermi level is well above the peaks in the potential well, resulting in a homogeneous 2D current flow for annealing and back-gating. When the carrier density is lowered to an intermediate level, some of the potential landscape is revealed, resulting in current holes in the 2D flow. The width of the back-gated potential has also started to shrink, but the features of the potential landscape are still preserved. Reducing the carrier density even further, the sample approached the transition to an insulator. The carriers are located in the narrow wells at the bottom of the potential, and the current flows along well-defined channels. The width reduction by the back-gate has increased to a point where features (i.e., the leftmost channel) are removed and the current redistributed toward the other channels. In the annealing case, the potential width is largely kept fixed with only small variations observed in Figure 2a, except for regions where large insulating patches develop as in Figure 3f.

**Discussion**

We now turn to discuss the implications of our results, comparing the evolution of local current flow as conductivity is reduced by electrostatic gating and by controlling the oxygen vacancy concentrations. Electrostatic back-gating is simple to operate and compatible with dynamic *in situ* changes of the carrier density in a range of measurement setups. However, back-gating changes the electronic properties beyond the carrier density. It alters the depth of the 2DES,[32] changes the mobility,[32] causes hysteresis,[30] affects the tetragonal domains,[3] and may cause leakage current. On the other hand, oxygen annealing presents a complementary approach, which changes the state of the sample in a quasi-permanent manner. By direct removal of donors from the crystal lattice, the full range from metallic to insulating is available, independent of the initial sheet carrier density.[21] Annealing also affects the sample mobility, similarly to back-gating, and is incompatible with tuning the carrier density in real-time at cryogenic temperatures, resulting in the need for thermally cycling the sample, which can change the microscopic current flow by altering the domain landscape.[16,36]

The images of the current flow as a function of annealing and gating reveal that the effects of the two approaches on the current distribution are very similar-the main difference being the increased effect of back-gating on the edge of the device. This can pose a problem if the features of the potential landscape are of interest, as for example the channels in our study are, as it is possible that features near the edge will be removed while those in the center will be hidden by the increase in carriers. However, this does present a potential combination of the two methods, where annealing is used to generate the initial resistive state, which is further modified at 4 K by a back-gate. In this study the back-gate could potentially be used in combination with annealing to investigate a transition from several conducting channels down to just one, thus forming an alternative approach where large devices defined by deposition through a shadow mask may be used to make quantum structures such as quantum point contacts without the need for lithography.[37,38]

The conducting channels observed here resemble the channels of modulated current due to domains observed in crystalline LAO/STO.[16,39] However, in crystalline LAO/STO the channels are typically visible in the as-deposited resistive state and become more dominant closer to the insulator,[18] while our observations of amorphous LAO/STO show channels that are generally revealed only closer to the insulator transition. It is possible that the oxygen vacancies are the cause of this difference between crystalline and amorphous top layers. This could be either through the vacancies themselves, as they can form linear clusters in STO,[40,41] or through an interaction with the walls separating different domains, where the formation energy of oxygen vacancies is lower than the bulk in certain ferroelectrics.[42] The exact interaction between vacancies and domains in a-LAO/STO is unclear, but our data indicate that domains should only gradually affect the current flow as the carrier density is reduced. The significance of the domains in modulating the current flow becomes more dramatic at smaller decides. At the nanoscale, a single domain could dominate the flow or even hinder it, depending on its orientation.

**Conclusion**

In this work, we studied the current distribution in a-LAO/STO while applying oxygen annealing and electrostatic gating to lower the carrier density nearing the insulator transition. We showed how annealing resulted in a transition from 2D homogeneous conduction to channel-like flow over several discrete channels. We compared this to current distributions tuned by electrostatic gating, where the sample width continuously shrinks. The observed changes to the current distribution have possible implications on the functionality and performance of mesoscopic and nanoscale devices. The observed transformation from 2D homogeneous current flow to discrete channels is desirable especially if control over the number of channels is achievable. Furthermore, information about the interplay between oxygen vacancies and the domain structure is of interest for domain engineering and complex oxides in general.

## Methods

*Materials Fabrication.* A-LAO top films were fabricated on TiO$_2$-terminated STO (001) substrate surfaces using pulsed laser deposition. Photoresist (AZ1505) was spin-coated on the surface followed by photolithography defining a resist mask for the geometry shown in Figure 1a. During deposition, the laser fluence was kept constant at 30 mJ with a 3 Hz repetition rate and oxygen partial pressure of $9 \times 10^{-7}$ mbar. The deposition temperature was kept constant at room temperature. These conditions result in a ~16 nm thick a-LAO layer. Before annealing and cryogenic measurements the samples were glued to an aluminum sample holder by using GE varnish with aluminum wire bonding for electrical connections.

*Scanning SQUID.* We combined the scanning SQUID setup with annealing capabilities by equipping the scanner with a 50 W heater, which we thermally anchored to the sample mount. They were then thermally decoupled from the system with a PCB spacing layer to protect the electronics and piezo positioners from unwanted temperature modulations. In this setup it was possible to heat the LAO/STO samples to 420 K in a controlled gas environment while measuring the transport properties and cool it down in liquid He to 4.2 K without exposing the sample to ambient conditions. During annealing and cooling the sheet resistance was measured in a four-probe setup with a DC current source of 1 µA. For scanning SQUID measurements an alternating current of 40-60 µA at 863 Hz was applied, and the generated magnetic flux was recorded by the SQUID using standard lock-in techniques. The annealed sample described in the text was also subjected to a back-gate voltage which changed the resistance but did not result in clear visual changes to the current distribution.

*Carrier Density Approximations.* While the measurement setup used here does not allow us to measure the carrier density from Hall measurements, we provide some estimates of the carrier density. A typical electron mobility for a-LAO/STO at room temperature is $\mu = 10 \frac{cm^2}{Vs}$ (refs [24,43]), from which the room temperature sheet carrier density can be extracted by using the relation $R_s = \frac{1}{en_s\mu}$. See Table S1 for a full reference of the transport data.

## ACKNOWLEDGMENT

We thank Eylon Persky and Naor Vardi for help with the experimental setup and the measurements. A.V.B. and B.K. were supported by the European Research Council Grant ERC-2019-COG-866236, Israeli Science Foundation Grant ISF-1281/17, the QuantERA ERA-NET Cofund in Quantum Technologies (Project 731473), and the Pazy Research Foundation. N.P. and D.V.C. were supported by the Villum Fonden, through the NEED project (00027993), the European Union's Horizon 2020, Future and Emerging Technologies (FET) program (Grant 801267), the Independent Research Fund Denmark (IRFD) – DFF-Research Project 3 (Thematic Research) - PIloT, Grant 0217-00069B, and the "Challenge Programme 2021 - Smart Nanomaterials for Applications in Life-Science" Grant NNF21OC0066526.

# Supporting Information for
# Current Mapping of Amorphous LaAlO$_3$/SrTiO$_3$ near the Metal-Insulator Transition

Anders V. Bjørlig,[1] Dennis V. Christensen,[2] Ricci Erlandsen,[2] Nini Pryds[2] and Beena Kalisky[1]*

[1]Department of Physics and Institute of Nanotechnology and Advanced Materials, Bar-Ilan University, Ramat Gan 5290002, Israel

[2]Department of Energy Conversion and Storage, Technical University of Denmark, Fysikvej 8 310, 2800 Kgs. Lyngby, Denmark

*Email: beena@biu.ac.il

## 1. Transport data for annealing

The annealing data described in this study came from measurements of an a-LAO/STO sample which was annealed 4 times. The calculated sheet densities and approximated carrier densities can be found in Table S1. See methods for the conversion to carrier density. In addition to this a more complete overview of the transport data for annealing step 1 can be seen in Figure S1.

| Annealing Step (#) | $R_S$ @ 290K (kΩ/■) | $n_s$ ($\cdot 10^{12}$ cm$^{-2}$) | Annealing Time @370 K (min) | Resistance rate @370 K (kΩ/■)/min |
|---|---|---|---|---|
| 0 | 88 | 7.1 | 13 | 0.93 |
| 1 | 100.5 | 6.2 | 18 | 1.25 |
| 2 | 120 | 5.2 | 35 | 1.39 |
| 3 | 156 | 4.0 | 20 | 3.7 |
| 4 | 300 | 2.1 | - | - |

**Table S1** – Sheet resistance, carrier density and annealing characteristics for each step.

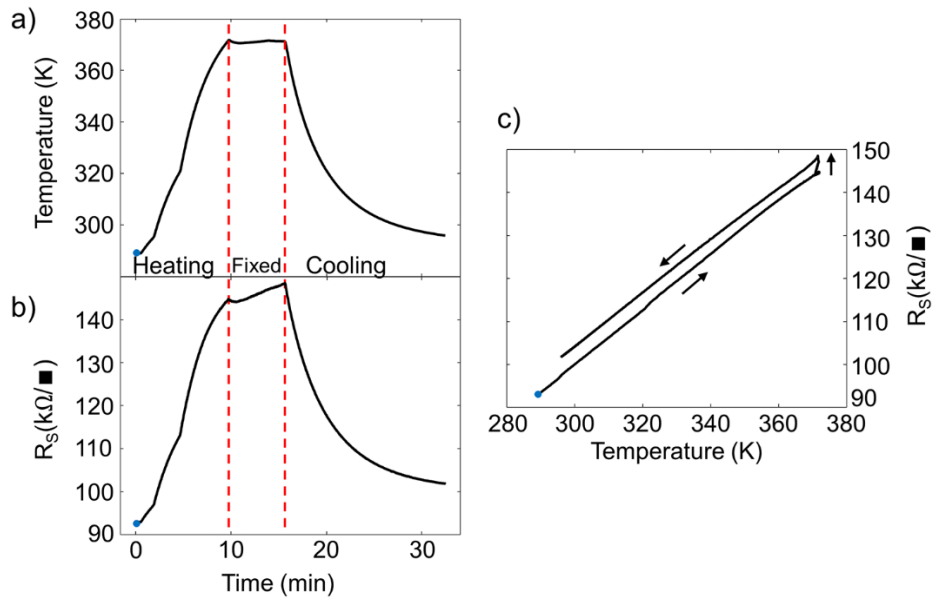

**Figure S1** – **Transport data overview for annealing step 1. a** Temperature recording during the annealing step where the sample is subjected to 1) heating, where current to the heater is gradually increased, 2) a fixed temperature, where the temperature is stabilized at 370 K, 3) cooling, where the heater is turned off and the sample returns to ambient temperature. **b** The corresponding changes to the sheet resistance during annealing. During warm-up and cool-down, the resistance changes due to a modification of the phonon scattering as well as an elimination of oxygen vacancies. When the temperature is fixed at 370 K, only the latter occurs. **c** The combined plot of resistance with temperature. The blue dot indicates the start of the measurement in all figures.

## 2. Thermal recovery in amorphous LAO/STO

When LAO/STO is subjected to a positive back-gate voltage at cryogenic temperatures some carriers will be lost to exposed charge traps.[1,2] This results in increased resistance which can be recovered by warming the sample to room temperature. During warm-up, previous studies reveal that the resistance show bumps around 65 K and 165 K, corresponding to specific energy barrier values.[1,3] Our data shows similar bumps at T = 77 K and 190 K (slightly higher than previously reported), but also 2 new signatures at T = 28 K and 240 K. In some cases, the resistance at room temperature did not recover to its pre cool-down value.

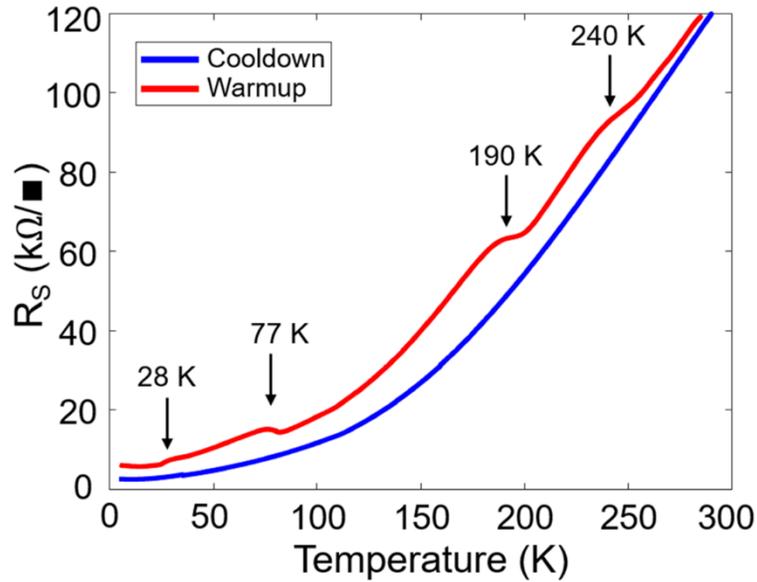

**Figure S2** – **Cool-down and warm-up resistance curves.** The initial cool-down curve (blue) shows a smooth decrease of resistance. After applying back-gate voltages from -150 V to 150 V at T = 4.2 K, the warm-up curve (red) has an increased resistance value and 4 bumps at T = 28, 77, 190 and 240 K (highlighted with arrows).

### 3. Formation of large current holes

We observe that large areas of reduced current density form and grow with each annealing step in some areas of the device (Figure S3). This significantly redistributes the current flow.

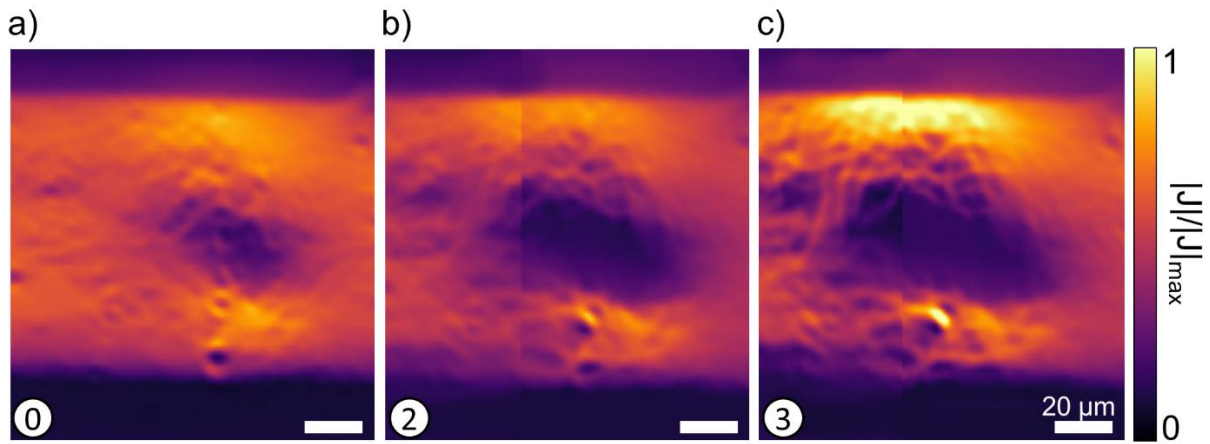

**Figure S3** – **Current distributions for cool-down 0,2 and 3 of the annealed sample. a** A large patch of reduced current density has formed, with some current still flowing inside the patch. **b** The patch has grown and the current is fully expelled on its inside. The edges of the patch are similar to the fractal current flow near the insulator transition observed with both annealing and gating. **c** The patch has grown even more and affects the full width of the channel.

## 4. Transport data during back-gate operation

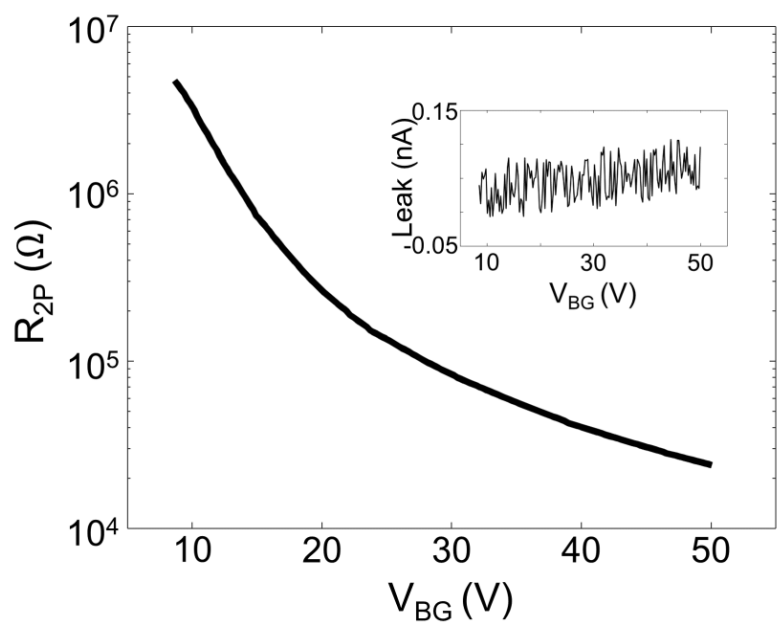

**Figure S4 – Transport data from back-gate operation.** The two-terminal resistance in response to positive back-gate voltage shows a smooth dependence. Inset: Leakage current was less than 0.2 nA during measurements.

## 5. Surface AFM images of the STO substrates

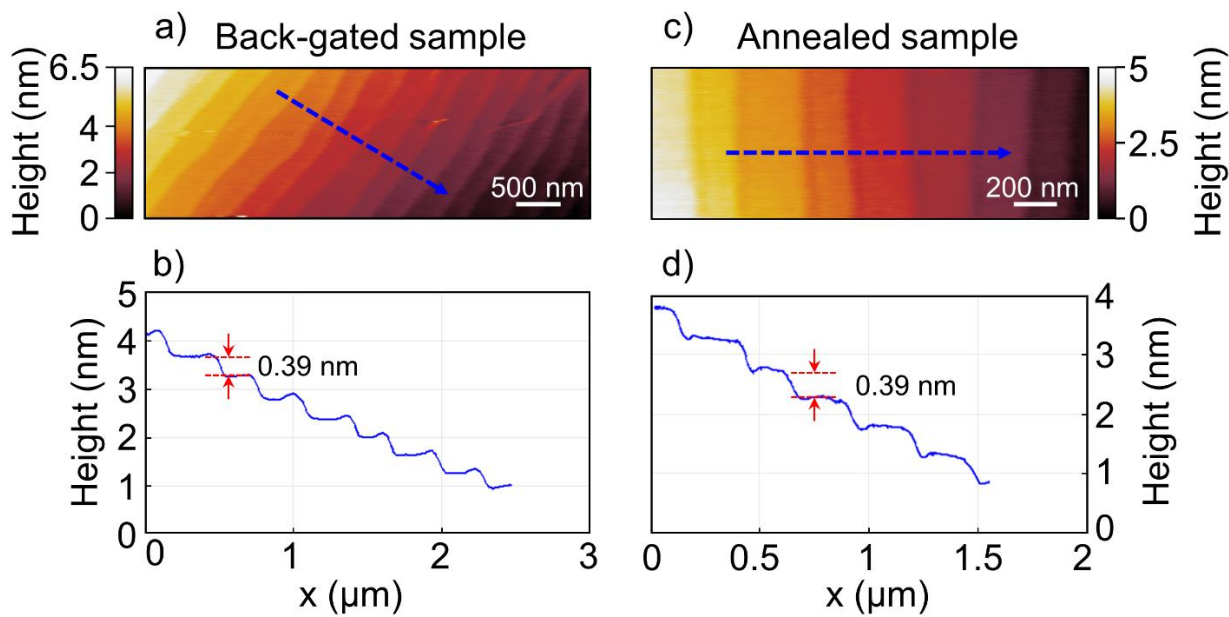

**Figure S5** – **AFM images of the STO substrates. a-b** AFM image of the Ti terminated (001) STO substrate of the back-gated sample revealing a clear terrace structure with the known 0.39 nm step height. **c-d** AFM images of the annealed sample with a similar structure to panel a-b.